\documentclass[aps,prl,twocolumn,showpacs,floatfix,nofootinbib,preprintnumbers,superscriptaddress,amsmath,amssymb]{revtex4-1}

\usepackage{graphicx}
\usepackage{epsfig}
\usepackage{bm}
\usepackage{color}
\usepackage{float}
\usepackage{dcolumn}
\usepackage{multirow}

\newcommand{\beq}{\begin{equation}}
\newcommand{\eeq}{\end{equation}}
\newcommand{\beqa}{\begin{eqnarray}}
\newcommand{\eeqa}{\end{eqnarray}}

\newcommand{\NNopt}{NNLO$_{\rm opt}$}
\newcommand{\NNEM}{N$^3$LO$_{\rm EM}$}

\begin{document}

\title{An optimized chiral nucleon-nucleon interaction at next-to-next-to-leading order}

\author{A.~Ekstr\"om} \affiliation{Department of Physics and Center of
  Mathematics for Applications, University of Oslo, N-0316 Oslo,
  Norway} \affiliation{National Superconducting Cyclotron Laboratory,
  Michigan State University, East Lansing, MI 48824, USA}

\author{G.~Baardsen} \affiliation{Department of Physics and Center of
  Mathematics for Applications, University of Oslo, N-0316 Oslo,
  Norway}

\author{C.~Forss\'en} \affiliation{Department of Fundamental Physics,
  Chalmers University of Technology, SE-412 96 G\"oteborg, Sweden}

\author{G.~Hagen} \affiliation{Physics Division, Oak Ridge National
  Laboratory, Oak Ridge, TN 37831, USA} 
  \affiliation{Department of
  Physics and Astronomy, University of Tennessee, Knoxville, TN 37996,
  USA} 

\author{M.~Hjorth-Jensen} \affiliation{Department of Physics
  and Center of Mathematics for Applications, University of Oslo,
  N-0316 Oslo, Norway} \affiliation{National Superconducting Cyclotron
  Laboratory, Michigan State University, East Lansing, MI 48824,
  USA} \affiliation{Department of Physics and Astronomy, Michigan
  State University, East Lansing, MI 48824, USA} 

\author{G.~R.~Jansen}
\affiliation{Physics Division, Oak Ridge National Laboratory, Oak
  Ridge, TN 37831, USA} \affiliation{Department of Physics and
  Astronomy, University of Tennessee, Knoxville, TN 37996, USA}

\author{R.~Machleidt} \affiliation{Department of
  Physics, University of Idaho, Moscow, ID 83844, USA}

\author{W.~Nazarewicz} \affiliation{Department of Physics and
  Astronomy, University of Tennessee, Knoxville, TN 37996, USA}
\affiliation{Physics Division, Oak Ridge National Laboratory, Oak
  Ridge, TN 37831, USA} 
\affiliation{Faculty of Physics, University of Warsaw, ul. Ho\.za 69, 00-681 Warsaw, Poland}  

\author{T.~Papenbrock} \affiliation{Department
  of Physics and Astronomy, University of Tennessee, Knoxville, TN
  37996, USA} \affiliation{Physics Division, Oak Ridge National
  Laboratory, Oak Ridge, TN 37831, USA} 

\author{J.~Sarich}
\affiliation{Mathematics and Computer Science Division, Argonne
  National Laboratory, Argonne, IL 60439, USA} 

\author{S.~M.~Wild}
\affiliation{Mathematics and Computer Science Division, Argonne
  National Laboratory, Argonne, IL 60439, USA}

\begin{abstract} 
  We optimize the nucleon-nucleon interaction from chiral effective
  field theory at next-to-next-to-leading order. The resulting new
  chiral force {\NNopt} yields $\chi^2\approx 1$ per degree of freedom
  for laboratory energies below approximately 125\,MeV. In the $A=3,4$
  nucleon systems, the contributions of three-nucleon forces are
  smaller than for previous parametrizations of chiral
  interactions. We use {\NNopt} to study properties of key nuclei and
  neutron matter, and we demonstrate that many aspects of nuclear
  structure can be understood in terms of this nucleon-nucleon
  interaction, without explicitly invoking three-nucleon forces.
\end{abstract}


\maketitle

{\it Introduction} -- Interactions from chiral effective field theory
(EFT) employ symmetries and the pattern of spontaneous symmetry
breaking of quantum chromodynamics~\cite{vankolck1994,EHM09,ME11}. In
this approach, the exchange of pions within chiral perturbation theory
yields the long-ranged contributions of the nuclear interaction, while
short-ranged components are included as contact terms. The interaction
is parametrized in terms of low-energy constants (LECs) that are
determined by fit to experimental data. The interactions from chiral
EFT exhibit a power counting in the ratio $Q/\Lambda$, with $Q$ being
the low-momentum scale being probed and $\Lambda$ the cutoff, which is
of the order of 1\,GeV. At next-to-next-to-leading order (NNLO),
three-nucleon forces (3NFs) enter, while four-nucleon forces (4NFs)
enter at next-to-next-to-next-to-leading order (N$^3$LO). For
laboratory energies below 125\,MeV, the nucleon-nucleon ($NN$) force
exhibits a quality of fit with $\chi^2\approx 10$/datum at NNLO
\cite{EGM04,*epelbaum2002a}, while a high-precision potential {\NNEM},
with a $\chi^2\approx 1$/datum up to 290\,MeV, was obtained by Entem
and Machleidt \cite{EM03,*ME11}.

The 3NFs at NNLO that accompany the current N$^3$LO $NN$ potentials
play a pivotal role in nuclear structure calculations
\cite{Hammer13}. They determine the ground-state spin of
$^{10}$B~\cite{Nav072}, correctly set the drip line in oxygen
isotopes~\cite{otsuka2010,hagen2012a}, and make $^{48}$Ca a doubly
magic nucleus~\cite{holt2012,hagen2012b}. While it might seem
surprising that smaller corrections at NNLO are so decisive for basic
nuclear structure properties, the 3NF contains spin-orbit and tensor
contributions that clearly are important for the currently employed
chiral interactions.  The contributions of 3NFs at N$^3$LO have also
been worked out~\cite{ishikawa2007,BEKM08,*BEKM11}, and there are
on-going efforts to compute even higher orders~\cite{krebs2012}.

While the quest for higher orders is important, this approach will
result in higher accuracy only if the optimization at lower orders
was carried out accurately. Thus, it is important and timely to
revisit the optimization question.  We note in particular that the
fits of the currently employed chiral
interactions~\cite{epelbaum2002a,epelbaum2002b,EM03} date back about a
decade and that there has been a considerable recent progress in
developing tools for the derivative-free nonlinear least-squares
optimization~\cite{Kor10}. Furthermore, the quantification of
theoretical uncertainties is a long-term objective of nuclear
structure theory, and this requires a covariance analysis of the
interaction parameters with respect to the experimental uncertainties
of the nucleon-nucleon elastic scattering observables; see, for example,
Refs.~\cite{reinhard2010,Kor10}. This letter takes the first step toward this
goal. We present a
state-of-the-art optimization of the $NN$ chiral EFT interaction at
NNLO. This yields a much-improved $\chi^2$ and a high-precision $NN$
potential {\NNopt}. The 3NF at NNLO is adjusted to the binding
energies in $A=3,4$ nuclei. We present computations of three-nucleon
and four-nucleon bound states, and we employ {\NNopt} to ground states and
excited states in $^{10}$B, masses and excited states of oxygen and
calcium isotopes, and neutron matter.

\emph{Optimizing the $NN$ interaction at NNLO} -- For the
optimization of the chiral $NN$ interaction we use the Practical
Optimization Using No Derivatives (for Squares) algorithm, 
POUNDerS~\cite{Kor10}, as implemented in ~\cite{tao-man}. This
derivative-free algorithm employs a quadratic model and is
particularly useful for computationally expensive objective functions.
We optimize the three pion-nucleon ($\pi N$) couplings ($c_1, c_3,
c_4$), and 11 partial wave contact parameters $C$ and $\tilde{C}$,
while we keep the axial-vector coupling constant $g_A$, the pion-decay
constant $f_{\pi}$, and all masses fixed.  In the optimization, we
minimize the objective function
\begin{equation}\label{chi2}
f(\vec{x}) = \sum_{q=1}^{N_q}
\left(\frac{\delta_q^{\textnormal{NNLO}}(\vec{x})
  -\delta_q^{\textnormal{Nijm93}}}{w_q}\right)^2,
\end{equation}
where $\delta^{\textnormal{NNLO}}$ are NNLO phase shifts,
$\delta^{\textnormal{Nijm93}}$ are experimental phase shifts from the
Nijmegen multi-energy partial-wave analysis~\cite{Sto93}, $\vec{x}$
denotes the parameters of the chiral interaction, and $w_q$ are
weighting factors. Note that Eq.~(\ref{chi2}) is not the $\chi^2$ with
respect to experimental data.  The actual $\chi^2$ is calculated
following the POUNDerS optimization. The phase shifts
$\delta^{\textnormal{NNLO}}$ are computed from $R$-matrix inversion,
and in the proton-proton ($pp$) channels we include the Coulomb
interaction~\cite{VP74,Lu94}.  The contact terms are optimized to
reproduce the Nijmegen phase shifts for each corresponding partial
wave, while keeping the $c_i$'s fixed. For the contacts, the weight
$w_q$ scales with the third power of the relative momentum $q$, while
for the $c_i$'s, we employ the uncertainties quoted in the Nijmegen
analysis~\cite{Sto93}. This approach 
can be justified by a physical argument:
for the peripheral waves the higher energies still represent
longer-range physics, and the need for a pedantic agreement with lower
energy phase shifts can be weakened.  The $\pi N$ couplings $c_1,c_3$,
and $c_4$ were simultaneously optimized to the peripheral
partial-waves ${}^1D_2,{}^3D_2,{}^3F_2,E_2,{}^3F_3,{}^1G_4,$ and ${}^3F_4$. Note
that the NNLO contact terms do not contribute to orbital angular
momenta $L\geq 2$.  We do not include other peripheral waves from the
Nijmegen study since they carry extremely small uncertainties, which
lead to a very noisy objective function.

Table~\ref{tab_nnlo_2body_parameters} summarizes the optimization
results.  Our values should be compared with the $\pi N$ couplings as
determined from $\pi N$ scattering data, where $c_1 = -0.81\pm0.15$,
$c_3=-4.69\pm1.34$, and $c_4=+3.40\pm0.04$ have been
obtained~\cite{BM00}.  Thus, POUNDerS yields values for $c_1$ and
$c_3$ that agree well with the empirical determination from $\pi N$
scattering. The $c_4$ value, however, deviates significantly from its
empirical value. The same trend was found in the construction of
the N$^3$LO~\cite{EM03} $NN$ interaction.  A detailed statistical
sensitivity analysis of the LECs with uncertainty quantification will
be presented in Ref.~\cite{tbp}.
%%%
\begin{table}[hbt]
\caption{\label{tab_nnlo_2body_parameters}
Parameters of  {\NNopt} at
  $\Lambda=500$ MeV and $\Lambda_{\textnormal{SFR}}=700$ MeV~\cite{EGM04}:  $c_i$ (in GeV$^{-1}$), $\tilde{C}$ (in $10^4$\,GeV$^{-2}$), and $C$ (in $10^4$\,GeV$^{-4}$).}
   \begin{ruledtabular} 
\begin{tabular}{cc|cc|cc}
LEC &  Value & LEC &  Value & LEC & Value \\ \hline   \\[-6pt]
$c_1$       & -0.91863953 & $c_3$ & -3.88868749 & $c_4$ &  4.31032716 \\
$\tilde{C}^{pp}_{{}^1S_0}$  & -0.15136604  & $\tilde{C}^{np}_{{}^1S_0}$  & -0.15214109  & $\tilde{C}^{nn}_{{}^1S_0}$  & -0.15176475 \\ 
$C_{{}^1S_0}$ &  2.40402194  & $C_{{}^3S_1}$ &  0.92838466 & $\tilde{C}_{{}^3S_1}$ & -0.15843418 \\
$C_{{}^1P_1}$ &  0.41704554  & $C_{{}^3P_0}$ &  1.26339076 & $C_{{}^3P_1}$ & -0.78265850 \\
$C_{{}^3S_1-{}^3D_1}$ &  0.61814142  & $C_{{}^3P_2}$  & -0.67780851  &&
\end{tabular}
  \end{ruledtabular}
\end{table}

Table~\ref{tab_nnlo_chi} shows the $\chi^2$/datum for {\NNopt} at
various laboratory energy bins. The quality of the fit is particularly
good for energies below 125\,MeV. For comparison, the $np$ NNLO
interaction of Ref.~\cite{EGM04} yields $\chi^2$/datum of 12--27 in
the range $\Lambda=600/700-450/500$~MeV at energies up to 290\,MeV.

\begin{table}[hbt]
\caption{\label{tab_nnlo_chi}
$\chi^2$/datum for  {\NNopt} at $\Lambda=500$~MeV and $\Lambda_{\textnormal{SFR}}=700$ MeV~\cite{EGM04} 
 with respect to the $np$ and $pp$ 1999
  databases~\cite{Mach01}. The values without the high-precision data sets
\cite{Cox67,*Jar71} are marked by asterisks.}
     \begin{ruledtabular} 
\begin{tabular}{c|cccc|c}
$T_{\rm lab}$ (MeV)& 0--35 & 35--125 & 125--183 & 183--290 & {\bf 0--290} \\  \hline        
$pp$ $\chi^2$/datum &  1.11 & 1.56    & $\left\{\begin{array}{c}23.95^{}\\4.35^*\end{array}\right.$ & 29.26 & $\left\{\begin{array}{c}{\bf 17.10}^{} \\ {\bf 14.03}^*\end{array}\right.$\\ 
$np$ $\chi^2$/datum &  0.85 & 1.17    & 1.87     & 6.09     & {\bf 2.95}
\end{tabular}
  \end{ruledtabular}
\end{table}
%%%%%
Around energies of 144~MeV there exist two data sets of $pp$
differential cross sections with a very high precision (0.5\%
error)~\cite{Cox67,*Jar71} (47 data points). The total number of $pp$
data in the energy interval 125--183~MeV is 343. The unusual precision
of those 47 data points distorts the $\chi^2$/datum for this
interval. For this reason, Table~\ref{tab_nnlo_chi} also shows the
results without the high-precision data.

Two comments are in order.  First, the $\chi^2$ with respect to
scattering observables is lower when the ${}^1P_1$ phase shifts are
weighted with the uncertainties from the Nijmegen analysis.  The
$P$-waves are accurately reproduced only when going to
N$^3$LO~\cite{EM03}. Second, the ${}^3S_1-{}^3D_1$ coupled channel is
optimized with the additional constraint of reproducing the deuteron
binding energy. The remaining deuteron observables, as well as the
${}^1S_0$ scattering observables, are predictions and reproduce the
experimental values well; see Table~\ref{tab_nnlo_observables}.
%%%%%%
\begin{table}[htb]
\caption{\label{tab_nnlo_observables}
 Scattering lengths $a$ and effective ranges $r$ (both in fm).
 The superscripts $N$ and $C$ for the proton-proton observables
  refer to  nuclear forces and Coulomb-plus-nuclear forces, respectively.
  $B_D$, $r_D$, $Q_D$, and $P_D$ denote the deuteron binding
  energy, radius, quadrupole moment, and
  $D$-state probability, respectively. $Q_D$ and $r_D$ are
  calculated without meson-exchange currents and relativistic
  corrections.}
     \begin{ruledtabular} 
\begin{tabular}{ccccl} 
 & {\NNEM} & {\NNopt} & Exp. & Ref. \\ \hline
\multirow{2}{*} {$a^C_{pp}$} & \multirow{2}{*}{-7.8188} & \multirow{2}{*}{-7.8174} & -7.8196(26) &\cite{Ber88} \\
          &         &         & -7.8149(29) &\cite{San83} \\
\multirow{2}{*}{$r^C_{pp}$} & \multirow{2}{*}{2.795}   &  \multirow{2}{*}{2.755}  &  2.790(14) &\cite{Ber88}  \\
          &         &         &  2.769(14) & \cite{San83} \\
$a^N_{pp}$ & -17.083 & -17.825 &  & \\ 
$r^N_{pp}$ & 2.876   & 2.817   & & \\
$a_{nn}$ & -18.900 & -18.889 & -18.95(40) & \cite{Gon06,Chen08} \\
$r_{nn}$ &  2.838  & 2.797   & 2.75(11) & \cite{Miller90} \\
$a_{np}$   & -23.732 & -23.749 & -23.740(20) & \cite{Mach01} \\
$r_{np}$   & 2.725   & 2.684   & 2.77(5) & \cite{Mach01} \\ [6pt] \hline
$B_D$ (MeV) & 2.224575 & 2.224582 & 2.224575(9) & \cite{Mach01} \\
$r_D$ (fm)  & 1.975 & 1.967 & 1.97535(85) & \cite{Huber98} \\
$Q_D$ (fm$^2$) & 0.275 & 0.272 & 0.2859(3) & \cite{Mach01} \\
$P_D$ (\%)        & 4.51 & 4.05 & &
\end{tabular}
   \end{ruledtabular} 
\end{table}

\begin{figure}[thb]
\includegraphics[width=0.9\columnwidth]{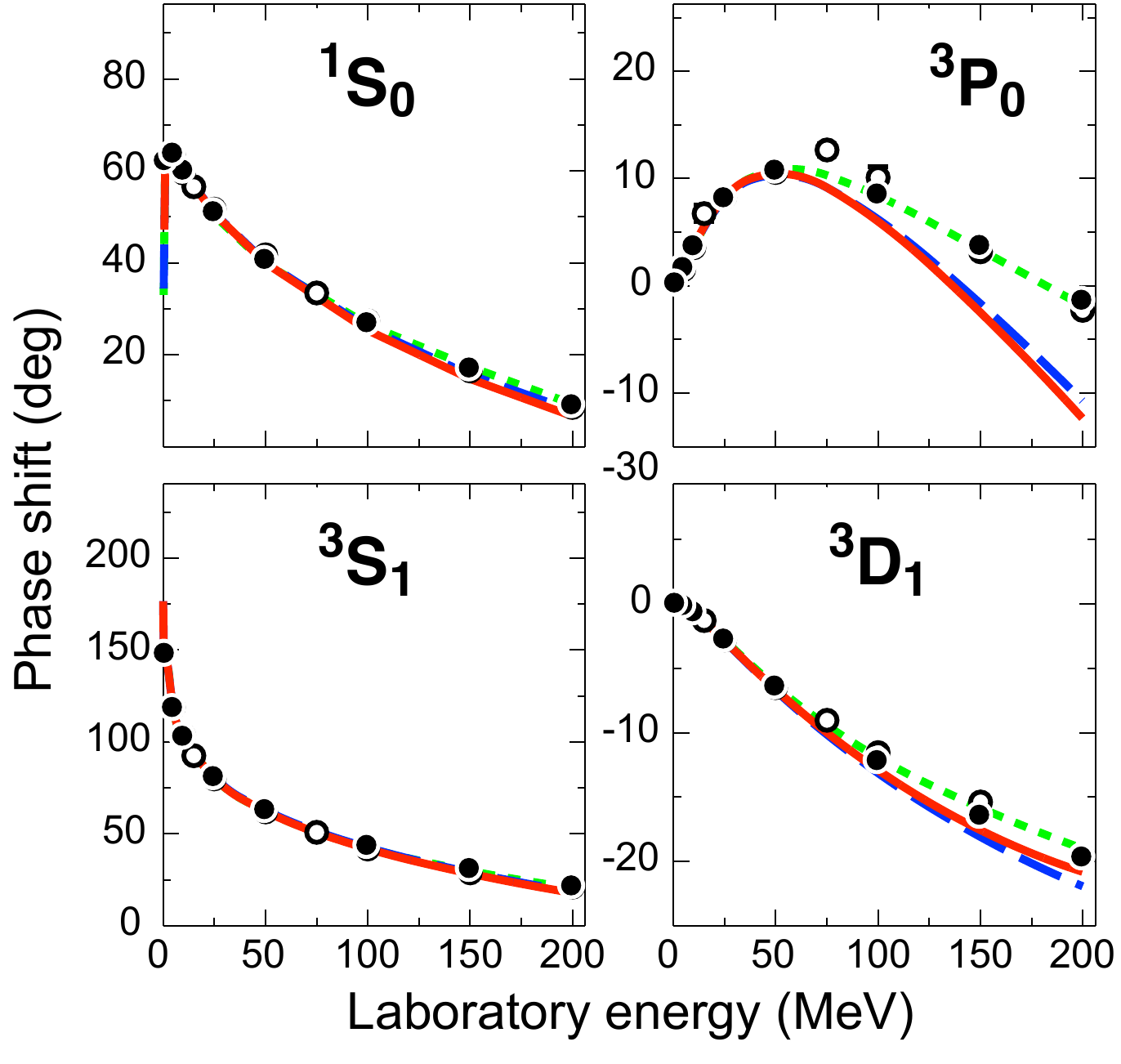}
\caption{(Color online) Computed $np$ phase shifts of the optimized
  NNLO potential of this work (red), the NNLO potential of
  Ref.~\cite{EGM04} (dashed, blue), and the N$^3$LO
  potential~\cite{EM03} (green, dotted) compared with the Nijmegen phase
  shift analysis~\protect\cite{Sto93} (solid dots) and the VPI/GWU
  analysis SM99~\cite{SM99} (open circles).}
\label{fig_phases1}
\end{figure}

Figure~\ref{fig_phases1} shows some $np$ phase shifts of {\NNopt} and
compares them with phase shifts from other potentials and partial wave
analyses.  Apart from the ${}^3P$-waves, the phase shifts of {\NNopt}
closely agree with those obtained at N$^3$LO. Note,
however, that these deviations do not spoil the good $\chi^2$ at
laboratory energies below 125~MeV.
%%%

%%%%%
\begin{table}[hbt]
  \caption{\label{tab_nnlo3NF_A34}
  Ground-state energies (in MeV) and point proton radii (in fm) 
    for $^3$H, $^3$He, and $^4$He
    using the {\NNopt} with and without the NNLO 3NF interaction
    for $c_D=-0.20$ and $c_E = -0.36$.}
       \begin{ruledtabular} 
\begin{tabular}{ccccc} 
            &      $E(^3$H)      & $E(^3$He)  & $E(^4$He)  & $r_p(^4$He) \\ \hline 
NNLO        &      -8.249     & -7.501  & -27.759 & 1.43(8) \\
NNLO+NNN    &      -8.469     & -7.722  & -28.417 & 1.43(8) \\
Experiment  &      -8.482     & -7.717  & -28.296 & 1.467(13) 
\end{tabular}
   \end{ruledtabular} 
\end{table}
%%%%
Three-nucleon forces also appear at NNLO, and two additional LECs
($c_D$ and $c_E$) enter. These are determined from calculations in the
three-nucleon and four-nucleon systems. We find that the binding
energies of $^3$H, $^3$He, and $^4$He do not uniquely determine $c_D$
and $c_E$, and the parametric dependence of both LECs is very similar
to those found in previous studies~\cite{Nog06,Nav072,Gaz09}.  Therefore, we
choose $c_D=-0.2$ guided by the triton
half life \cite{Gaz09} and obtain $c_E=-0.36$ from optimization to the
binding energies. The resulting point charge radii of $^4$He are also
in good agreement with experiment; see Table~\ref{tab_nnlo3NF_A34}.

\emph{Performance of {\rm \NNopt} for light- and medium-mass nuclei and
  neutron matter} -- In this paper, we apply {\NNopt} to $^{10}$B,
isotopes of oxygen and calcium, and neutron matter. The considered
systems are particularly interesting because the current $NN$ chiral
interactions at N$^3$LO completely fail to describe key aspects of
their structure.

To study the ground- and first excited state in $^{10}$B, we carry out
no-core shell model (configuration interaction)
calculations~\cite{Barrett:2013-69} using the bare {\NNopt} in model
spaces of up to $N_{\rm max}=10$ harmonic oscillator (HO) shells
(10\,$\hbar\Omega$) above the unperturbed configuration. These model
spaces are not large enough to provide fully converged results for the
ground state and first excited state of $^{10}$B. Still, the variational
upper bounds for the energies are $-54.35$~MeV for the $1^+$ state and
$-54.32$~MeV for the $3^+$ state.  The energies are very close, in
contrast to {\NNEM}, which yields a level spacing of about 1.2~MeV
between the $J^\pi=1^+$ ground state and the $J^\pi=3^+$ excited
state~\cite{Nav072}.

\begin{figure}[hbtp]
    \includegraphics[width=0.9\columnwidth]{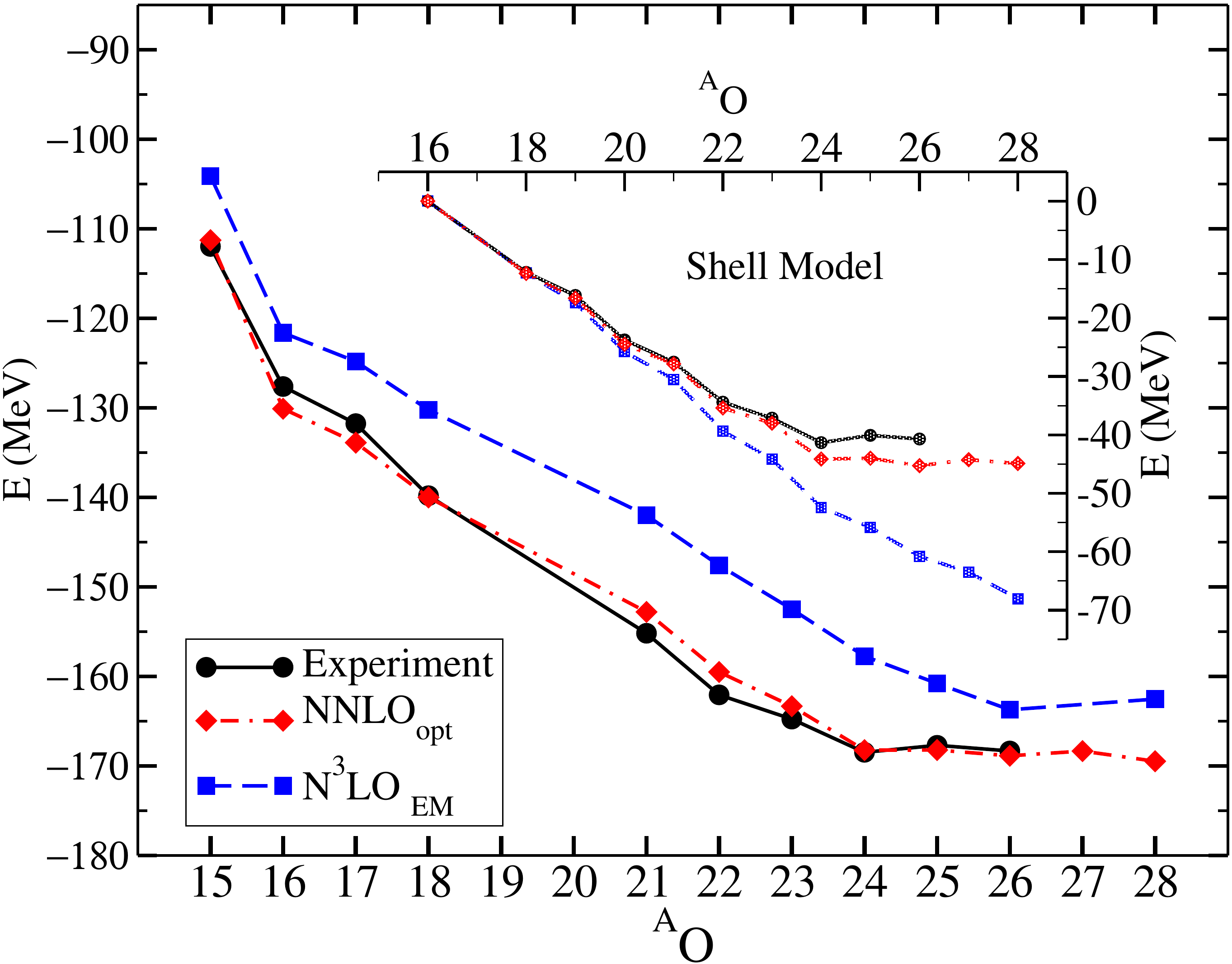}
  \caption{(Color online). The ground-state energies of oxygen
    isotopes obtained in CC with the {\NNopt} and {\NNEM} interactions
    compared with experiment. The inset shows SM results.}
  \label{fig:ccOx}
\end{figure}

Chiral $NN$ interactions at N$^3$LO fail to explain the neutron
drip-line in oxygen isotopes, and 3NFs have been the key element for 
understanding the structure of nuclei around
$^{24}$O~\cite{otsuka2010,hagen2012a}.  Figure~\ref{fig:ccOx} shows
the experimental ground-state energies of oxygen isotopes and
compares the results from coupled-cluster (CC) computations in the
$\Lambda$ triples
approximation~\cite{kucharski1998,taube2008,hagen2010a}. Our CC
calculations employ a Hartree-Fock basis (HF) built from $N_{\rm
  max}=15$ HO shells at $\hbar\Omega = 20$\,MeV. Because of the
``softness'' of {\NNopt}, this model space is sufficiently large to
converge the ground states and excited states of the nuclei considered.  In
addition, we performed shell-model (SM) calculations assuming the
closed $^{16}$O core with an effective interaction derived from
many-body perturbation theory to third order in the interaction and
including folded diagrams~\cite{hko1995}. For the SM calculations, the
single-particle energies were taken from the experimental $^{17}$O
spectrum.  In both CC and SM, {\NNopt} results are close to
experiment. In contrast, the {\NNEM} case requires
3NFs to provide reasonable description of measured values.
%%%%%%
%%%%%%%

\begin{figure}[hbtp]
    \includegraphics[width=0.9\columnwidth]{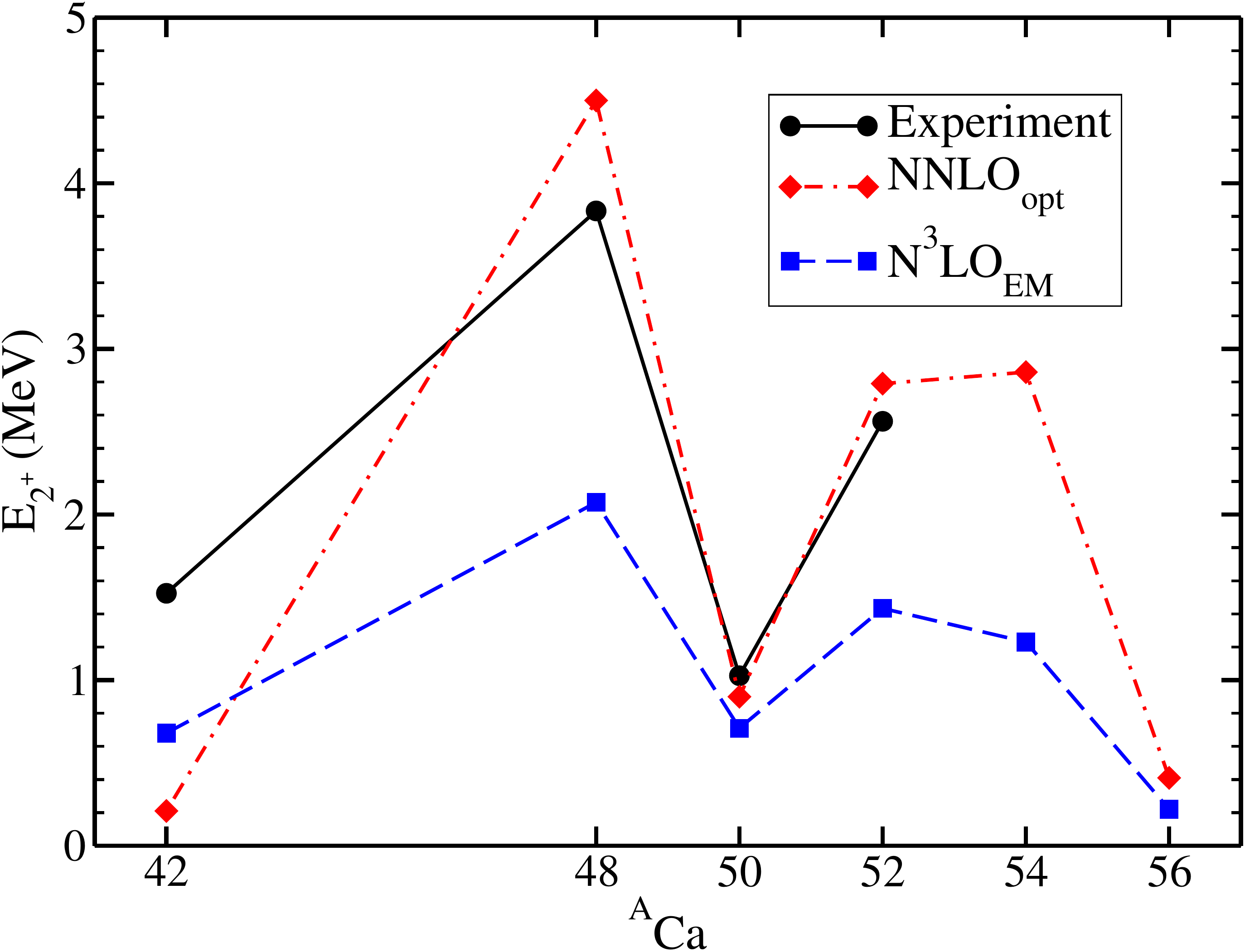}
  \caption{(Color online). The first $2_1^+$ state in selected calcium isotopes 
  obtained in CC with the {\NNopt} and {\NNEM} interactions compared with 
experiment.}
  \label{fig:ca2plus}
\end{figure}

Now we consider the heavy isotopes of calcium. Here, $^{48}$Ca is
doubly magic, $^{52}$Ca exhibits a soft subshell closure, and
$^{54}$Ca is predicted to have an even softer subshell
closure~\cite{hagen2012b}. A signature of shell closure is the
location of the first $2^+$ state. We employed CC equation-of-motion
methods within the singles and doubles approximation
\cite{hagen2010a,jansen2011} to compute the first $2^+$ state in the
calcium isotopes. Figure~\ref{fig:ca2plus} shows that {\NNEM} fails to
describe the location of the first $2^+$ state in
$^{40,48,50,52,54,56}$Ca. In contrast, {\NNopt} yields $^{48}$Ca as a
doubly magic nucleus and predicts subshell closures in $^{52,54}$Ca.
%%%%
%%%%%
The {\NNopt} overbinds the calcium isotopes by about 1\,MeV per
nucleon. In particular $^{40,48,52}$Ca are overbound by 1.03\,MeV,
1.06\,MeV, and 1.04\,MeV per nucleon, respectively. That is, the excess
energy per nucleon is fairly constant; hence, {\NNopt} reproduces
binding energy differences, such as neutron-separation energies and
low-lying excited states, rather well.

The complete description of nuclei at NNLO also requires 3NFs.  We
computed the first $2^+$ state in $^{22,24}$O and in $^{48}$Ca with
the 3NF compatible with the \NNopt interaction. The matrix elements of
the 3NF are expensive computationally, and we must at present 
limit their calculation to three-body energies up to $e_{3{\rm max}} =
2n_a+l_a + 2n_b+l_b +2n_c+l_c=14$.  (Recall that we employ $15$ major
harmonic oscillator shells for the $NN$ interaction.) We also used the
normal ordered two-body approximation for the 3NF
\cite{hagen2007,binder2013a} with respect to a HF reference.  With the
restriction of $e_{3{\rm max}} = 14$, we were not able to obtain fully
converged results for the binding energies of oxygen and calcium
isotopes.  However, excitation energies relative to the ground state
converge somewhat better. Our results for the first $2^+$ state in
$^{22,24}$O and in $^{48}$Ca are 2.3(3)\,MeV, 3.5(5)\,MeV and
4.8(7)\,MeV, respectively. We estimate the uncertainty by varying
$\hbar\Omega$ in the interval 16--22\,MeV. The results obtained by using
{\NNopt} $NN$ interaction alone yields 2.5\,MeV, 5.0\,MeV, and
4.5\,MeV in $^{22,24}$O and $^{48}$Ca, respectively.  These
preliminary results suggest that the 3NFs may not dramatically change
the results that were obtained with the {\NNopt} $NN$ interaction
alone.

\begin{figure}[hbtp]
    \includegraphics[width=0.9\columnwidth,clip=]{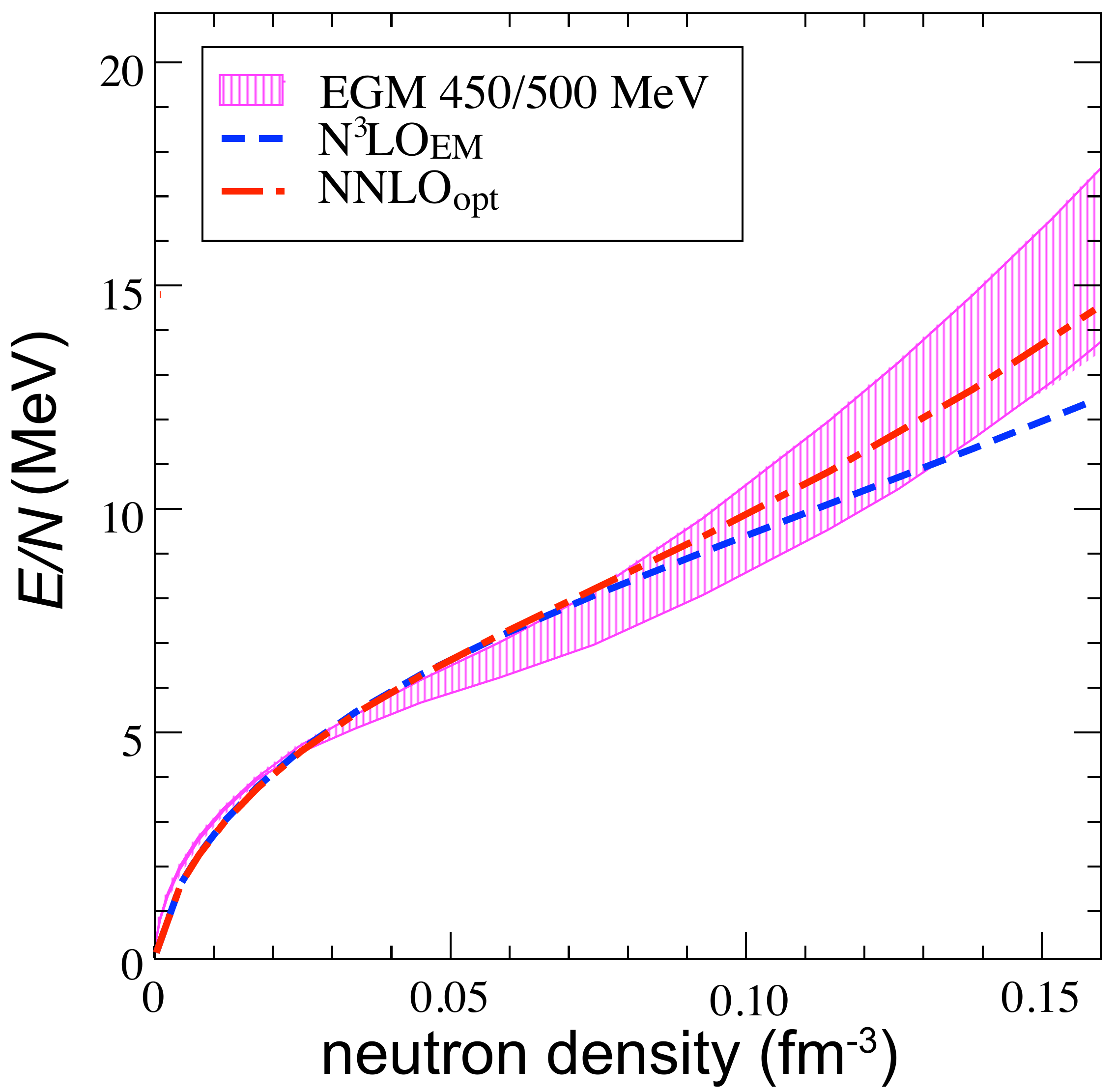}
  \caption{(Color online). Energy per nucleon for neutron matter for
    {\NNopt} and {\NNEM} \cite{EM03}. The calculations used the CC
    method with the inclusion of particle-particle ladders and a
    continuous single-particle spectrum. The shaded area (EGM) shows
    uncertainty bands for N$^3$LO chiral effective field theory
    calculations of Ref.~\cite{Tews2013}, including 3NFs.  }
  \label{fig:neutronmatter}
\end{figure}

It is instructive to compare the predictions of {\NNopt} and {\NNEM}
for the neutron matter equation of state at sub-saturation densities
with the results of ab-initio calculations of Refs.~\cite{Tews2013}.
Figure~\ref{fig:neutronmatter} shows that the performance of {\NNopt}
is on par with the EGM results of Ref.~\cite{Tews2013}, which take
into account the effects of 3NFs and 4NFs.  The predictions of {\NNEM}
deviate from other results at higher densities.
%%%%% 

{\it Conclusions} -- We constructed the new $NN$ chiral EFT
interaction {\NNopt} at next-to-next-to-leading order using the
optimization tool POUNDerS in the phase-shift analysis.  The
optimization of the low-energy constants in the $NN$-sector at NNLO
yields a $\chi^2$/datum of about one for laboratory scattering
energies below 125~MeV. The {\NNopt} $NN$ interaction yields very
good agreement with binding energies and radii for $A=3,4$ nuclei. Key
aspects of nuclear structure, such as excitation spectra, the position
of the neutron drip line in oxygen, shell-closures in calcium, and the
neutron matter equation of state at sub-saturation densities, are
reproduced by {\NNopt} interaction alone, without resorting to
3NFs. We performed the initial calculation of the first $2^+$ states
in $^{22,24}$O and $^{48}$Ca with {\NNopt} supplemented by a 3NF and
found effects of 3NFs to be small and good agreement with
experimental excitation energies.  The precise role of 3NFs in
medium-mass nuclei, the quantification of theoretical uncertainties,
and optimizations at higher-order chiral interactions will be
addressed in forthcoming investigations.

\begin{acknowledgments}
  We thank M.~P.~ Kartamyshev, B.~D.~Carlsson, and H.~T.~Johansson for
  discussions and related code development. This work was supported by
  the Research Council of Norway under contract ISP-Fysikk/216699; by
  the Office of Nuclear Physics, U.S. Department of Energy (Oak Ridge
  National Laboratory), under Grant Nos.~DE-FG02-03ER41270 (University
  of Idaho), DE-FG02-96ER40963 (University of Tennessee), DE-AC02-06CH11357
  (Argonne), and
  DE-SC0008499 (NUCLEI SciDAC collaboration); by the Swedish Research
  Council (dnr 2007-4078), and by the European Research Council
  (ERC-StG-240603). Computer time was provided by the Innovative and
  Novel Computational Impact on Theory and Experiment (INCITE)
  program. This research used resources of the Oak Ridge Leadership
  Computing Facility located in the Oak Ridge National Laboratory, which
  is supported by the Office of Science of the Department of Energy
  under Contract No. DE-AC05-00OR22725, and used computational resources
  of the National Center for Computational Sciences, the National
  Institute for Computational Sciences, and the Notur project in Norway.
\end{acknowledgments}

\bibliography{n2lo_letter}

\begin{thebibliography}{46}
\expandafter\ifx\csname natexlab\endcsname\relax\def\natexlab#1{#1}\fi
\expandafter\ifx\csname bibnamefont\endcsname\relax
  \def\bibnamefont#1{#1}\fi
\expandafter\ifx\csname bibfnamefont\endcsname\relax
  \def\bibfnamefont#1{#1}\fi
\expandafter\ifx\csname citenamefont\endcsname\relax
  \def\citenamefont#1{#1}\fi
\expandafter\ifx\csname url\endcsname\relax
  \def\url#1{\texttt{#1}}\fi
\expandafter\ifx\csname urlprefix\endcsname\relax\def\urlprefix{URL }\fi
\providecommand{\bibinfo}[2]{#2}
\providecommand{\eprint}[2][]{\url{#2}}

\bibitem[{\citenamefont{van Kolck}(1994)}]{vankolck1994}
\bibinfo{author}{\bibfnamefont{U.}~\bibnamefont{van Kolck}},
  \bibinfo{journal}{Phys. Rev. C} \textbf{\bibinfo{volume}{49}},
  \bibinfo{pages}{2932} (\bibinfo{year}{1994}).

\bibitem[{\citenamefont{Epelbaum et~al.}(2009)\citenamefont{Epelbaum, Hammer,
  and Mei\ss{}ner}}]{EHM09}
\bibinfo{author}{\bibfnamefont{E.}~\bibnamefont{Epelbaum}},
  \bibinfo{author}{\bibfnamefont{H.-W.} \bibnamefont{Hammer}},
  \bibnamefont{and} \bibinfo{author}{\bibfnamefont{U.-G.}
  \bibnamefont{Mei\ss{}ner}}, \bibinfo{journal}{Rev. Mod. Phys.}
  \textbf{\bibinfo{volume}{81}}, \bibinfo{pages}{1773} (\bibinfo{year}{2009}).

\bibitem[{\citenamefont{Machleidt and Entem}(2011)}]{ME11}
\bibinfo{author}{\bibfnamefont{R.}~\bibnamefont{Machleidt}} \bibnamefont{and}
  \bibinfo{author}{\bibfnamefont{D.}~\bibnamefont{Entem}},
  \bibinfo{journal}{Phys. Rep.} \textbf{\bibinfo{volume}{503}},
  \bibinfo{pages}{1 } (\bibinfo{year}{2011}).

\bibitem[{\citenamefont{Epelbaum et~al.}(2004)\citenamefont{Epelbaum,
  Gl\"ockle, and Mei\ss{}ner}}]{EGM04}
\bibinfo{author}{\bibfnamefont{E.}~\bibnamefont{Epelbaum}},
  \bibinfo{author}{\bibfnamefont{W.}~\bibnamefont{Gl\"ockle}},
  \bibnamefont{and} \bibinfo{author}{\bibfnamefont{U.-G.}
  \bibnamefont{Mei\ss{}ner}}, \bibinfo{journal}{Eur. Phys. J. A}
  \textbf{\bibinfo{volume}{19}}, \bibinfo{pages}{401} (\bibinfo{year}{2004}).

\bibitem[{\citenamefont{Epelbaum
  et~al.}(2002{\natexlab{a}})\citenamefont{Epelbaum, Nogga, Gl{\"o}ckle,
  Kamada, Mei{\ss}ner, and Wita{\l}a}}]{epelbaum2002a}
\bibinfo{author}{\bibfnamefont{E.}~\bibnamefont{Epelbaum}},
  \bibinfo{author}{\bibfnamefont{A.}~\bibnamefont{Nogga}},
  \bibinfo{author}{\bibfnamefont{W.}~\bibnamefont{Gl{\"o}ckle}},
  \bibinfo{author}{\bibfnamefont{H.}~\bibnamefont{Kamada}},
  \bibinfo{author}{\bibfnamefont{U.-G.} \bibnamefont{Mei{\ss}ner}},
  \bibnamefont{and}
  \bibinfo{author}{\bibfnamefont{H.}~\bibnamefont{Wita{\l}a}},
  \bibinfo{journal}{Eur. Phys. J. A} \textbf{\bibinfo{volume}{15}},
  \bibinfo{pages}{543} (\bibinfo{year}{2002}{\natexlab{a}}), ISSN
  \bibinfo{issn}{1434-6001}.

\bibitem[{\citenamefont{Entem and Machleidt}(2003)}]{EM03}
\bibinfo{author}{\bibfnamefont{D.~R.} \bibnamefont{Entem}} \bibnamefont{and}
  \bibinfo{author}{\bibfnamefont{R.}~\bibnamefont{Machleidt}},
  \bibinfo{journal}{Phys. Rev. C} \textbf{\bibinfo{volume}{68}},
  \bibinfo{pages}{041001(R)} (\bibinfo{year}{2003}).

\bibitem[{\citenamefont{Hammer et~al.}(2013)\citenamefont{Hammer, Nogga, and
  Schwenk}}]{Hammer13}
\bibinfo{author}{\bibfnamefont{H.-W.} \bibnamefont{Hammer}},
  \bibinfo{author}{\bibfnamefont{A.}~\bibnamefont{Nogga}}, \bibnamefont{and}
  \bibinfo{author}{\bibfnamefont{A.}~\bibnamefont{Schwenk}},
  \bibinfo{journal}{Rev. Mod. Phys.} \textbf{\bibinfo{volume}{85}},
  \bibinfo{pages}{197} (\bibinfo{year}{2013}).

\bibitem[{\citenamefont{Navr\'atil et~al.}(2007)\citenamefont{Navr\'atil,
  Gueorguiev, Vary, Ormand, and Nogga}}]{Nav072}
\bibinfo{author}{\bibfnamefont{P.}~\bibnamefont{Navr\'atil}},
  \bibinfo{author}{\bibfnamefont{V.~G.} \bibnamefont{Gueorguiev}},
  \bibinfo{author}{\bibfnamefont{J.~P.} \bibnamefont{Vary}},
  \bibinfo{author}{\bibfnamefont{W.~E.} \bibnamefont{Ormand}},
  \bibnamefont{and} \bibinfo{author}{\bibfnamefont{A.}~\bibnamefont{Nogga}},
  \bibinfo{journal}{Phys. Rev. Lett.} \textbf{\bibinfo{volume}{99}},
  \bibinfo{pages}{042501} (\bibinfo{year}{2007}).

\bibitem[{\citenamefont{Otsuka et~al.}(2010)\citenamefont{Otsuka, Suzuki, Holt,
  Schwenk, and Akaishi}}]{otsuka2010}
\bibinfo{author}{\bibfnamefont{T.}~\bibnamefont{Otsuka}},
  \bibinfo{author}{\bibfnamefont{T.}~\bibnamefont{Suzuki}},
  \bibinfo{author}{\bibfnamefont{J.~D.} \bibnamefont{Holt}},
  \bibinfo{author}{\bibfnamefont{A.}~\bibnamefont{Schwenk}}, \bibnamefont{and}
  \bibinfo{author}{\bibfnamefont{Y.}~\bibnamefont{Akaishi}},
  \bibinfo{journal}{Phys. Rev. Lett.} \textbf{\bibinfo{volume}{105}},
  \bibinfo{pages}{032501} (\bibinfo{year}{2010}).

\bibitem[{\citenamefont{Hagen et~al.}(2012{\natexlab{a}})\citenamefont{Hagen,
  Hjorth-Jensen, Jansen, Machleidt, and Papenbrock}}]{hagen2012a}
\bibinfo{author}{\bibfnamefont{G.}~\bibnamefont{Hagen}},
  \bibinfo{author}{\bibfnamefont{M.}~\bibnamefont{Hjorth-Jensen}},
  \bibinfo{author}{\bibfnamefont{G.~R.} \bibnamefont{Jansen}},
  \bibinfo{author}{\bibfnamefont{R.}~\bibnamefont{Machleidt}},
  \bibnamefont{and}
  \bibinfo{author}{\bibfnamefont{T.}~\bibnamefont{Papenbrock}},
  \bibinfo{journal}{Phys. Rev. Lett.} \textbf{\bibinfo{volume}{108}},
  \bibinfo{pages}{242501} (\bibinfo{year}{2012}{\natexlab{a}}).

\bibitem[{\citenamefont{Holt et~al.}(2012)\citenamefont{Holt, Otsuka, Schwenk,
  and Suzuki}}]{holt2012}
\bibinfo{author}{\bibfnamefont{J.~D.} \bibnamefont{Holt}},
  \bibinfo{author}{\bibfnamefont{T.}~\bibnamefont{Otsuka}},
  \bibinfo{author}{\bibfnamefont{A.}~\bibnamefont{Schwenk}}, \bibnamefont{and}
  \bibinfo{author}{\bibfnamefont{T.}~\bibnamefont{Suzuki}},
  \bibinfo{journal}{J. Phys. G} \textbf{\bibinfo{volume}{39}},
  \bibinfo{pages}{085111} (\bibinfo{year}{2012}).

\bibitem[{\citenamefont{Hagen et~al.}(2012{\natexlab{b}})\citenamefont{Hagen,
  Hjorth-Jensen, Jansen, Machleidt, and Papenbrock}}]{hagen2012b}
\bibinfo{author}{\bibfnamefont{G.}~\bibnamefont{Hagen}},
  \bibinfo{author}{\bibfnamefont{M.}~\bibnamefont{Hjorth-Jensen}},
  \bibinfo{author}{\bibfnamefont{G.~R.} \bibnamefont{Jansen}},
  \bibinfo{author}{\bibfnamefont{R.}~\bibnamefont{Machleidt}},
  \bibnamefont{and}
  \bibinfo{author}{\bibfnamefont{T.}~\bibnamefont{Papenbrock}},
  \bibinfo{journal}{Phys. Rev. Lett.} \textbf{\bibinfo{volume}{109}},
  \bibinfo{pages}{032502} (\bibinfo{year}{2012}{\natexlab{b}}).

\bibitem[{\citenamefont{Ishikawa and Robilotta}(2007)}]{ishikawa2007}
\bibinfo{author}{\bibfnamefont{S.}~\bibnamefont{Ishikawa}} \bibnamefont{and}
  \bibinfo{author}{\bibfnamefont{M.~R.} \bibnamefont{Robilotta}},
  \bibinfo{journal}{Phys. Rev. C} \textbf{\bibinfo{volume}{76}},
  \bibinfo{pages}{014006} (\bibinfo{year}{2007}).

\bibitem[{\citenamefont{Bernard et~al.}(2008)\citenamefont{Bernard, Epelbaum,
  Krebs, and Mei\ss{}ner}}]{BEKM08}
\bibinfo{author}{\bibfnamefont{V.}~\bibnamefont{Bernard}},
  \bibinfo{author}{\bibfnamefont{E.}~\bibnamefont{Epelbaum}},
  \bibinfo{author}{\bibfnamefont{H.}~\bibnamefont{Krebs}}, \bibnamefont{and}
  \bibinfo{author}{\bibfnamefont{U.-G.} \bibnamefont{Mei\ss{}ner}},
  \bibinfo{journal}{Phys. Rev. C} \textbf{\bibinfo{volume}{77}},
  \bibinfo{pages}{064004} (\bibinfo{year}{2008}).

\bibitem[{\citenamefont{Bernard et~al.}(2011)\citenamefont{Bernard, Epelbaum,
  Krebs, and Mei\ss{}ner}}]{BEKM11}
\bibinfo{author}{\bibfnamefont{V.}~\bibnamefont{Bernard}},
  \bibinfo{author}{\bibfnamefont{E.}~\bibnamefont{Epelbaum}},
  \bibinfo{author}{\bibfnamefont{H.}~\bibnamefont{Krebs}}, \bibnamefont{and}
  \bibinfo{author}{\bibfnamefont{U.-G.} \bibnamefont{Mei\ss{}ner}},
  \bibinfo{journal}{Phys. Rev. C} \textbf{\bibinfo{volume}{84}},
  \bibinfo{pages}{054001} (\bibinfo{year}{2011}).

\bibitem[{\citenamefont{Krebs et~al.}(2012)\citenamefont{Krebs, Gasparyan, and
  Epelbaum}}]{krebs2012}
\bibinfo{author}{\bibfnamefont{H.}~\bibnamefont{Krebs}},
  \bibinfo{author}{\bibfnamefont{A.}~\bibnamefont{Gasparyan}},
  \bibnamefont{and} \bibinfo{author}{\bibfnamefont{E.}~\bibnamefont{Epelbaum}},
  \bibinfo{journal}{Phys. Rev. C} \textbf{\bibinfo{volume}{85}},
  \bibinfo{pages}{054006} (\bibinfo{year}{2012}).

\bibitem[{\citenamefont{Epelbaum
  et~al.}(2002{\natexlab{b}})\citenamefont{Epelbaum, Nogga, Gl\"ockle, Kamada,
  Mei\ss{}ner, and Wita{\l}a}}]{epelbaum2002b}
\bibinfo{author}{\bibfnamefont{E.}~\bibnamefont{Epelbaum}},
  \bibinfo{author}{\bibfnamefont{A.}~\bibnamefont{Nogga}},
  \bibinfo{author}{\bibfnamefont{W.}~\bibnamefont{Gl\"ockle}},
  \bibinfo{author}{\bibfnamefont{H.}~\bibnamefont{Kamada}},
  \bibinfo{author}{\bibfnamefont{U.-G.} \bibnamefont{Mei\ss{}ner}},
  \bibnamefont{and}
  \bibinfo{author}{\bibfnamefont{H.}~\bibnamefont{Wita{\l}a}},
  \bibinfo{journal}{Phys. Rev. C} \textbf{\bibinfo{volume}{66}},
  \bibinfo{pages}{064001} (\bibinfo{year}{2002}{\natexlab{b}}).

\bibitem[{\citenamefont{Kortelainen et~al.}(2010)\citenamefont{Kortelainen,
  Lesinski, Mor\'e, Nazarewicz, Sarich, Schunck, Stoitsov, and Wild}}]{Kor10}
\bibinfo{author}{\bibfnamefont{M.}~\bibnamefont{Kortelainen}},
  \bibinfo{author}{\bibfnamefont{T.}~\bibnamefont{Lesinski}},
  \bibinfo{author}{\bibfnamefont{J.}~\bibnamefont{Mor\'e}},
  \bibinfo{author}{\bibfnamefont{W.}~\bibnamefont{Nazarewicz}},
  \bibinfo{author}{\bibfnamefont{J.}~\bibnamefont{Sarich}},
  \bibinfo{author}{\bibfnamefont{N.}~\bibnamefont{Schunck}},
  \bibinfo{author}{\bibfnamefont{M.~V.} \bibnamefont{Stoitsov}},
  \bibnamefont{and} \bibinfo{author}{\bibfnamefont{S.}~\bibnamefont{Wild}},
  \bibinfo{journal}{Phys. Rev. C} \textbf{\bibinfo{volume}{82}},
  \bibinfo{pages}{024313} (\bibinfo{year}{2010}).

\bibitem[{\citenamefont{Reinhard and Nazarewicz}(2010)}]{reinhard2010}
\bibinfo{author}{\bibfnamefont{P.-G.} \bibnamefont{Reinhard}} \bibnamefont{and}
  \bibinfo{author}{\bibfnamefont{W.}~\bibnamefont{Nazarewicz}},
  \bibinfo{journal}{Phys. Rev. C} \textbf{\bibinfo{volume}{81}},
  \bibinfo{pages}{051303} (\bibinfo{year}{2010}).

\bibitem[{\citenamefont{Munson et~al.}(2012)\citenamefont{Munson, Sarich, Wild,
  Benson, and {Curfman McInnes}}}]{tao-man}
\bibinfo{author}{\bibfnamefont{T.}~\bibnamefont{Munson}},
  \bibinfo{author}{\bibfnamefont{J.}~\bibnamefont{Sarich}},
  \bibinfo{author}{\bibfnamefont{S.~M.} \bibnamefont{Wild}},
  \bibinfo{author}{\bibfnamefont{S.}~\bibnamefont{Benson}}, \bibnamefont{and}
  \bibinfo{author}{\bibfnamefont{L.}~\bibnamefont{{Curfman McInnes}}},
  \bibinfo{type}{Technical Memorandum} \bibinfo{number}{ANL/MCS-TM-322},
  \bibinfo{institution}{Argonne National Laboratory},
  \bibinfo{address}{Argonne, Illinois} (\bibinfo{year}{2012}),
  \bibinfo{note}{see \url{http://www.mcs.anl.gov/tao}}.

\bibitem[{\citenamefont{Stoks et~al.}(1993)\citenamefont{Stoks, Klomp,
  Rentmeester, and de~Swart}}]{Sto93}
\bibinfo{author}{\bibfnamefont{V.~G.~J.} \bibnamefont{Stoks}},
  \bibinfo{author}{\bibfnamefont{R.~A.~M.} \bibnamefont{Klomp}},
  \bibinfo{author}{\bibfnamefont{M.~C.~M.} \bibnamefont{Rentmeester}},
  \bibnamefont{and} \bibinfo{author}{\bibfnamefont{J.~J.}
  \bibnamefont{de~Swart}}, \bibinfo{journal}{Phys. Rev. C}
  \textbf{\bibinfo{volume}{48}}, \bibinfo{pages}{792} (\bibinfo{year}{1993}).

\bibitem[{\citenamefont{Vincent and Phatak}(1974)}]{VP74}
\bibinfo{author}{\bibfnamefont{C.~M.} \bibnamefont{Vincent}} \bibnamefont{and}
  \bibinfo{author}{\bibfnamefont{S.~C.} \bibnamefont{Phatak}},
  \bibinfo{journal}{Phys. Rev. C} \textbf{\bibinfo{volume}{10}},
  \bibinfo{pages}{391} (\bibinfo{year}{1974}).

\bibitem[{\citenamefont{Lu et~al.}(1994)\citenamefont{Lu, Mefford, Landau, and
  Song}}]{Lu94}
\bibinfo{author}{\bibfnamefont{D.~H.} \bibnamefont{Lu}},
  \bibinfo{author}{\bibfnamefont{T.}~\bibnamefont{Mefford}},
  \bibinfo{author}{\bibfnamefont{R.~H.} \bibnamefont{Landau}},
  \bibnamefont{and} \bibinfo{author}{\bibfnamefont{G.}~\bibnamefont{Song}},
  \bibinfo{journal}{Phys. Rev. C} \textbf{\bibinfo{volume}{50}},
  \bibinfo{pages}{3037} (\bibinfo{year}{1994}).

\bibitem[{\citenamefont{B\"uttiker and Mei\ss{}ner}(2000)}]{BM00}
\bibinfo{author}{\bibfnamefont{P.}~\bibnamefont{B\"uttiker}} \bibnamefont{and}
  \bibinfo{author}{\bibfnamefont{U.-G.} \bibnamefont{Mei\ss{}ner}},
  \bibinfo{journal}{Nuclear Physics A} \textbf{\bibinfo{volume}{668}},
  \bibinfo{pages}{97 } (\bibinfo{year}{2000}).

\bibitem[{\citenamefont{Ekstr{\"o}m et~al.}(2013)}]{tbp}
\bibinfo{author}{\bibfnamefont{A.}~\bibnamefont{Ekstr{\"o}m}}
  \bibnamefont{et~al.}, \bibinfo{howpublished}{to be published}
  (\bibinfo{year}{2013}).

\bibitem[{\citenamefont{Machleidt}(2001)}]{Mach01}
\bibinfo{author}{\bibfnamefont{R.}~\bibnamefont{Machleidt}},
  \bibinfo{journal}{Phys. Rev. C} \textbf{\bibinfo{volume}{63}},
  \bibinfo{pages}{024001} (\bibinfo{year}{2001}).

\bibitem[{\citenamefont{Cox et~al.}(1967)\citenamefont{Cox, Eaton, Zyl, Jarvis,
  and Rose}}]{Cox67}
\bibinfo{author}{\bibfnamefont{G.}~\bibnamefont{Cox}},
  \bibinfo{author}{\bibfnamefont{G.}~\bibnamefont{Eaton}},
  \bibinfo{author}{\bibfnamefont{C.~V.} \bibnamefont{Zyl}},
  \bibinfo{author}{\bibfnamefont{O.}~\bibnamefont{Jarvis}}, \bibnamefont{and}
  \bibinfo{author}{\bibfnamefont{B.}~\bibnamefont{Rose}},
  \bibinfo{journal}{Nuclear Physics B} \textbf{\bibinfo{volume}{4}},
  \bibinfo{pages}{353 } (\bibinfo{year}{1967}), \bibinfo{note}{21 pp diff.
  cross section data at 144.1 MeV}.

\bibitem[{\citenamefont{Jarvis et~al.}(1971)\citenamefont{Jarvis, Whitehead,
  and Shah}}]{Jar71}
\bibinfo{author}{\bibfnamefont{O.}~\bibnamefont{Jarvis}},
  \bibinfo{author}{\bibfnamefont{C.}~\bibnamefont{Whitehead}},
  \bibnamefont{and} \bibinfo{author}{\bibfnamefont{M.}~\bibnamefont{Shah}},
  \bibinfo{journal}{Physics Letters B} \textbf{\bibinfo{volume}{36}},
  \bibinfo{pages}{409 } (\bibinfo{year}{1971}), \bibinfo{note}{26 pp diff.
  cross section data at 144.0 MeV}.

\bibitem[{\citenamefont{Bergervoet et~al.}(1988)\citenamefont{Bergervoet, van
  Campen, van~der Sanden, and de~Swart}}]{Ber88}
\bibinfo{author}{\bibfnamefont{J.~R.} \bibnamefont{Bergervoet}},
  \bibinfo{author}{\bibfnamefont{P.~C.} \bibnamefont{van Campen}},
  \bibinfo{author}{\bibfnamefont{W.~A.} \bibnamefont{van~der Sanden}},
  \bibnamefont{and} \bibinfo{author}{\bibfnamefont{J.~J.}
  \bibnamefont{de~Swart}}, \bibinfo{journal}{Phys. Rev. C}
  \textbf{\bibinfo{volume}{38}}, \bibinfo{pages}{15} (\bibinfo{year}{1988}).

\bibitem[{\citenamefont{van~der Sanden et~al.}(1983)\citenamefont{van~der
  Sanden, Emmen, and de~Swart}}]{San83}
\bibinfo{author}{\bibfnamefont{W.~A.} \bibnamefont{van~der Sanden}},
  \bibinfo{author}{\bibfnamefont{A.~H.} \bibnamefont{Emmen}}, \bibnamefont{and}
  \bibinfo{author}{\bibfnamefont{J.~J.} \bibnamefont{de~Swart}},
  \bibinfo{type}{Tech. Rep.}, \bibinfo{institution}{Nijmegen}
  (\bibinfo{year}{1983}), \bibinfo{note}{unpublished}.

\bibitem[{\citenamefont{Gonzalez~Trotter
  et~al.}(2006)\citenamefont{Gonzalez~Trotter, Meneses, Tornow, Howell, Chen,
  Crowell, Roper, Walter, Schmidt, Wita\l{}a et~al.}}]{Gon06}
\bibinfo{author}{\bibfnamefont{D.~E.} \bibnamefont{Gonzalez~Trotter}},
  \bibinfo{author}{\bibfnamefont{F.~S.} \bibnamefont{Meneses}},
  \bibinfo{author}{\bibfnamefont{W.}~\bibnamefont{Tornow}},
  \bibinfo{author}{\bibfnamefont{C.~R.} \bibnamefont{Howell}},
  \bibinfo{author}{\bibfnamefont{Q.}~\bibnamefont{Chen}},
  \bibinfo{author}{\bibfnamefont{A.~S.} \bibnamefont{Crowell}},
  \bibinfo{author}{\bibfnamefont{C.~D.} \bibnamefont{Roper}},
  \bibinfo{author}{\bibfnamefont{R.~L.} \bibnamefont{Walter}},
  \bibinfo{author}{\bibfnamefont{D.}~\bibnamefont{Schmidt}},
  \bibinfo{author}{\bibfnamefont{H.}~\bibnamefont{Wita\l{}a}},
  \bibnamefont{et~al.}, \bibinfo{journal}{Phys. Rev. C}
  \textbf{\bibinfo{volume}{73}}, \bibinfo{pages}{034001}
  (\bibinfo{year}{2006}).

\bibitem[{\citenamefont{Chen et~al.}(2008)\citenamefont{Chen, Howell, Carman,
  Gibbs, Gibson, Hussein, Kiser, Mertens, Moore, Morris et~al.}}]{Chen08}
\bibinfo{author}{\bibfnamefont{Q.}~\bibnamefont{Chen}},
  \bibinfo{author}{\bibfnamefont{C.~R.} \bibnamefont{Howell}},
  \bibinfo{author}{\bibfnamefont{T.~S.} \bibnamefont{Carman}},
  \bibinfo{author}{\bibfnamefont{W.~R.} \bibnamefont{Gibbs}},
  \bibinfo{author}{\bibfnamefont{B.~F.} \bibnamefont{Gibson}},
  \bibinfo{author}{\bibfnamefont{A.}~\bibnamefont{Hussein}},
  \bibinfo{author}{\bibfnamefont{M.~R.} \bibnamefont{Kiser}},
  \bibinfo{author}{\bibfnamefont{G.}~\bibnamefont{Mertens}},
  \bibinfo{author}{\bibfnamefont{C.~F.} \bibnamefont{Moore}},
  \bibinfo{author}{\bibfnamefont{C.}~\bibnamefont{Morris}},
  \bibnamefont{et~al.}, \bibinfo{journal}{Phys. Rev. C}
  \textbf{\bibinfo{volume}{77}}, \bibinfo{pages}{054002}
  (\bibinfo{year}{2008}).

\bibitem[{\citenamefont{Miller et~al.}(1990)\citenamefont{Miller, Nefkens, and
  \ifmmode~\check{S}\else \v{S}\fi{}laus}}]{Miller90}
\bibinfo{author}{\bibfnamefont{G.}~\bibnamefont{Miller}},
  \bibinfo{author}{\bibfnamefont{B.}~\bibnamefont{Nefkens}}, \bibnamefont{and}
  \bibinfo{author}{\bibfnamefont{I.}~\bibnamefont{\ifmmode~\check{S}\else
  \v{S}\fi{}laus}}, \bibinfo{journal}{Phys. Rep.}
  \textbf{\bibinfo{volume}{194}}, \bibinfo{pages}{1 } (\bibinfo{year}{1990}).

\bibitem[{\citenamefont{Huber et~al.}(1998)\citenamefont{Huber, Udem, Gross,
  Reichert, Kourogi, Pachucki, Weitz, and H\"ansch}}]{Huber98}
\bibinfo{author}{\bibfnamefont{A.}~\bibnamefont{Huber}},
  \bibinfo{author}{\bibfnamefont{T.}~\bibnamefont{Udem}},
  \bibinfo{author}{\bibfnamefont{B.}~\bibnamefont{Gross}},
  \bibinfo{author}{\bibfnamefont{J.}~\bibnamefont{Reichert}},
  \bibinfo{author}{\bibfnamefont{M.}~\bibnamefont{Kourogi}},
  \bibinfo{author}{\bibfnamefont{K.}~\bibnamefont{Pachucki}},
  \bibinfo{author}{\bibfnamefont{M.}~\bibnamefont{Weitz}}, \bibnamefont{and}
  \bibinfo{author}{\bibfnamefont{T.~W.} \bibnamefont{H\"ansch}},
  \bibinfo{journal}{Phys. Rev. Lett.} \textbf{\bibinfo{volume}{80}},
  \bibinfo{pages}{468} (\bibinfo{year}{1998}).

\bibitem[{\citenamefont{Arndt et~al.}(1999)\citenamefont{Arndt, Strakovsky, and
  Workman}}]{SM99}
\bibinfo{author}{\bibfnamefont{R.~A.} \bibnamefont{Arndt}},
  \bibinfo{author}{\bibfnamefont{I.~I.} \bibnamefont{Strakovsky}},
  \bibnamefont{and} \bibinfo{author}{\bibfnamefont{R.~L.}
  \bibnamefont{Workman}} (\bibinfo{year}{1999}), \bibinfo{note}{{S}AID,
  Scattering Analysis Interactive Dial-in computer facility, George Washington
  University (formerly Virginia Polytechnic Institute), solution SM99 (Summer
  1999); for more information see, e.~g., R. A. Arndt, I. I. Strakovsky, and R.
  L. Workman, Phys. Rev. C {\bf 50}, 2731 (1994).}

\bibitem[{\citenamefont{Nogga et~al.}(2006)\citenamefont{Nogga, Navr\'atil,
  Barrett, and Vary}}]{Nog06}
\bibinfo{author}{\bibfnamefont{A.}~\bibnamefont{Nogga}},
  \bibinfo{author}{\bibfnamefont{P.}~\bibnamefont{Navr\'atil}},
  \bibinfo{author}{\bibfnamefont{B.~R.} \bibnamefont{Barrett}},
  \bibnamefont{and} \bibinfo{author}{\bibfnamefont{J.~P.} \bibnamefont{Vary}},
  \bibinfo{journal}{Phys. Rev. C} \textbf{\bibinfo{volume}{73}},
  \bibinfo{pages}{064002} (\bibinfo{year}{2006}).

\bibitem[{\citenamefont{Gazit et~al.}(2009)\citenamefont{Gazit, Quaglioni, and
  Navr\'atil}}]{Gaz09}
\bibinfo{author}{\bibfnamefont{D.}~\bibnamefont{Gazit}},
  \bibinfo{author}{\bibfnamefont{S.}~\bibnamefont{Quaglioni}},
  \bibnamefont{and}
  \bibinfo{author}{\bibfnamefont{P.}~\bibnamefont{Navr\'atil}},
  \bibinfo{journal}{Phys. Rev. Lett.} \textbf{\bibinfo{volume}{103}},
  \bibinfo{pages}{102502} (\bibinfo{year}{2009}).

\bibitem[{\citenamefont{Barrett et~al.}(2013)\citenamefont{Barrett, Navr\'atil,
  and Vary}}]{Barrett:2013-69}
\bibinfo{author}{\bibfnamefont{B.}~\bibnamefont{Barrett}},
  \bibinfo{author}{\bibfnamefont{P.}~\bibnamefont{Navr\'atil}},
  \bibnamefont{and} \bibinfo{author}{\bibfnamefont{J.~P.} \bibnamefont{Vary}},
  \bibinfo{journal}{Prog. Part. Nucl. Phys.} \textbf{\bibinfo{volume}{69}},
  \bibinfo{pages}{131} (\bibinfo{year}{2013}).

\bibitem[{\citenamefont{Kucharski and Bartlett}(1998)}]{kucharski1998}
\bibinfo{author}{\bibfnamefont{S.~A.} \bibnamefont{Kucharski}}
  \bibnamefont{and} \bibinfo{author}{\bibfnamefont{R.~J.}
  \bibnamefont{Bartlett}}, \bibinfo{journal}{The J.~Chem.~Phys.}
  \textbf{\bibinfo{volume}{108}}, \bibinfo{pages}{5243} (\bibinfo{year}{1998}).

\bibitem[{\citenamefont{Taube and Bartlett}(2008)}]{taube2008}
\bibinfo{author}{\bibfnamefont{A.~G.} \bibnamefont{Taube}} \bibnamefont{and}
  \bibinfo{author}{\bibfnamefont{R.~J.} \bibnamefont{Bartlett}},
  \bibinfo{journal}{The J.~Chem.~Phys.} \textbf{\bibinfo{volume}{128}},
  \bibinfo{pages}{044110} (\bibinfo{year}{2008}).

\bibitem[{\citenamefont{Hagen et~al.}(2010)\citenamefont{Hagen, Papenbrock,
  Dean, and Hjorth-Jensen}}]{hagen2010a}
\bibinfo{author}{\bibfnamefont{G.}~\bibnamefont{Hagen}},
  \bibinfo{author}{\bibfnamefont{T.}~\bibnamefont{Papenbrock}},
  \bibinfo{author}{\bibfnamefont{D.~J.} \bibnamefont{Dean}}, \bibnamefont{and}
  \bibinfo{author}{\bibfnamefont{M.}~\bibnamefont{Hjorth-Jensen}},
  \bibinfo{journal}{Phys. Rev. C} \textbf{\bibinfo{volume}{82}},
  \bibinfo{pages}{034330} (\bibinfo{year}{2010}).

\bibitem[{\citenamefont{Hjorth-Jensen et~al.}(1995)\citenamefont{Hjorth-Jensen,
  Kuo, and Osnes}}]{hko1995}
\bibinfo{author}{\bibfnamefont{M.}~\bibnamefont{Hjorth-Jensen}},
  \bibinfo{author}{\bibfnamefont{T.~T.~S.} \bibnamefont{Kuo}},
  \bibnamefont{and} \bibinfo{author}{\bibfnamefont{E.}~\bibnamefont{Osnes}},
  \bibinfo{journal}{Phys. Rep.} \textbf{\bibinfo{volume}{261}},
  \bibinfo{pages}{125} (\bibinfo{year}{1995}).

\bibitem[{\citenamefont{Jansen et~al.}(2011)\citenamefont{Jansen,
  Hjorth-Jensen, Hagen, and Papenbrock}}]{jansen2011}
\bibinfo{author}{\bibfnamefont{G.~R.} \bibnamefont{Jansen}},
  \bibinfo{author}{\bibfnamefont{M.}~\bibnamefont{Hjorth-Jensen}},
  \bibinfo{author}{\bibfnamefont{G.}~\bibnamefont{Hagen}}, \bibnamefont{and}
  \bibinfo{author}{\bibfnamefont{T.}~\bibnamefont{Papenbrock}},
  \bibinfo{journal}{Phys. Rev. C} \textbf{\bibinfo{volume}{83}},
  \bibinfo{pages}{054306} (\bibinfo{year}{2011}).

\bibitem[{\citenamefont{Hagen et~al.}(2007)\citenamefont{Hagen, Papenbrock,
  Dean, Schwenk, Nogga, W\l{}och, and Piecuch}}]{hagen2007}
\bibinfo{author}{\bibfnamefont{G.}~\bibnamefont{Hagen}},
  \bibinfo{author}{\bibfnamefont{T.}~\bibnamefont{Papenbrock}},
  \bibinfo{author}{\bibfnamefont{D.~J.} \bibnamefont{Dean}},
  \bibinfo{author}{\bibfnamefont{A.}~\bibnamefont{Schwenk}},
  \bibinfo{author}{\bibfnamefont{A.}~\bibnamefont{Nogga}},
  \bibinfo{author}{\bibfnamefont{M.}~\bibnamefont{W\l{}och}}, \bibnamefont{and}
  \bibinfo{author}{\bibfnamefont{P.}~\bibnamefont{Piecuch}},
  \bibinfo{journal}{Phys. Rev. C} \textbf{\bibinfo{volume}{76}},
  \bibinfo{pages}{034302} (\bibinfo{year}{2007}).

\bibitem[{\citenamefont{Binder et~al.}(2013)\citenamefont{Binder, Langhammer,
  Calci, Navr\'atil, and Roth}}]{binder2013a}
\bibinfo{author}{\bibfnamefont{S.}~\bibnamefont{Binder}},
  \bibinfo{author}{\bibfnamefont{J.}~\bibnamefont{Langhammer}},
  \bibinfo{author}{\bibfnamefont{A.}~\bibnamefont{Calci}},
  \bibinfo{author}{\bibfnamefont{P.}~\bibnamefont{Navr\'atil}},
  \bibnamefont{and} \bibinfo{author}{\bibfnamefont{R.}~\bibnamefont{Roth}},
  \bibinfo{journal}{Phys. Rev. C} \textbf{\bibinfo{volume}{87}},
  \bibinfo{pages}{021303} (\bibinfo{year}{2013}).

\bibitem[{\citenamefont{Tews et~al.}(2013)\citenamefont{Tews, Kr\"uger,
  Hebeler, and Schwenk}}]{Tews2013}
\bibinfo{author}{\bibfnamefont{I.}~\bibnamefont{Tews}},
  \bibinfo{author}{\bibfnamefont{T.}~\bibnamefont{Kr\"uger}},
  \bibinfo{author}{\bibfnamefont{K.}~\bibnamefont{Hebeler}}, \bibnamefont{and}
  \bibinfo{author}{\bibfnamefont{A.}~\bibnamefont{Schwenk}},
  \bibinfo{journal}{Phys. Rev. Lett.} \textbf{\bibinfo{volume}{110}},
  \bibinfo{pages}{032504} (\bibinfo{year}{2013}).

\end{thebibliography}
\bibliographystyle{apsrev}

\end{document}